# Investigation of Differential Diffusion and Strain Coupling in Large Eddy Simulations of Hydrogen-Air Flames

Antonio Masucci[a], Gioele Ferrante[b], Tiziano Ghisu[a], Andrea Giusti[c] and Ivan Langella[b]

[a] Department of Mechanical, Chemical and Materials Engineering, Università degli Studi di Cagliari, Cagliari, IT;
[b] Faculty of Aerospace Engineering, Delft University of Technology, Delft, NL
[c] Department of Mechanical Engineering, Imperial College, London, UK

**ABSTRACT**
Large Eddy Simulations with flamelet-based thermochemistry are used to investigate the behaviour of a premixed hydrogen-air flame stabilised by a bluff-body. Validation against experimental data is carried out first to demonstrate the model's ability to predict both velocity field and flame structure. The capability of the model in predicting differential diffusion effects is then assessed, in particular regarding the coupling between differential diffusion, tangential strain and curvature, and their effect on mixture fraction redistribution and reaction rate variation. Results indicate that unstretched flamelet thermochemistry is capable of capturing the increase in mixture fraction caused by positive resolved strain, as well as negative variations of mixture fraction due to negative curvature. Furthermore, the model is observed to mimic the effects of negative Markstein length to a certain extent, so that positive tangential strain causes reaction rate increase. The interplay between resolved stretch and preferential diffusion is also shown to lead to a shorter flame length which is in better agreement with experimental observations as compared to simulations under unity Lewis number assumption. These findings highlight that the macroscopic effects of differential diffusion and stretch on the premixed hydrogen flame, characterised by significant strain levels, can be predicted using a flamelet-based approach and without recurring to strained flamelets database, which implies important simplifications in the combustion modelling of turbulent hydrogen-premixed flames and offers valuable insights for the design of novel combustors.

**KEYWORDS**
Hydrogen; Bluff-body; Turbulence; Strain; Premixed Combustion

CONTACT A. Masucci. Email: antonio.masucci@unica.it

# 1. Introduction

Hydrogen combustion is carbon-free, offering significant potential for decarbonization in view of fighting the climate change. However, burning hydrogen poses challenges due to its distinct physical and chemical properties as compared to hydrocarbon fuels. Hydrogen exhibits high molecular diffusivity, low activation energy, and a high burning velocity which increase the risk of flashback. Moreover, hydrogen's flame high temperature may imply higher emissions of nitrogen oxides (NOx). To limit these, hydrogen can be burnt at ultra-lean conditions, implying lower temperatures and a reduction of NOx formation. However, utra-lean conditions are difficult to design due to hydrogen's preferential diffusion, whose effects and their modelling are still not fully understood. Large Eddy Simulation (LES) is a powerful computational tool for capturing unsteady combustion physics in practical configurations at an affordable computational cost as compared, for example, to direct numerical simulations (DNS). In an LES, only the large turbulent flow structures are resolved, with models used to mimic the effects of the small scales on the large motion. Since combustion is unduly a small scale phenomenon, the turbulence-flame interaction must be modelled in an LES.

Among the most common models for this interaction in premixed flames, flamelet-based approaches have been extensively used in the context of LES for hydrocarbon flames due to their good compromise between limited computational cost and relatively good accuracy. Nevertheless, flamelet models were mainly developed under the assumption of species equidiffusivity, and are thus unable to capture the effects of differential diffusion and its coupling with stretch which, in hydrogen flames, plays a crucial role on the flame dynamics even at moderate strain levels (Law 1989). In the context of flamelet approaches, while one-dimensional effects such as mixture leaning ahead of the flame can be accounted in the LES by introducing corrections to the mixture fraction, progress variable, and enthalpy equations, addressing two-dimensional effects of strain and curvature present greater challenges (Mukundakumar et al. 2021; Regele et al. 2013; Lapenna et al. 2021). In recent years, research effort has been oriented to the development of canonical combustion systems to gain understanding of the hydrogen combustion physics and the associated model development. Some examples include the lifted partially premixed flame in hot coflow (Cabra et al. 2002), the low swirl burner studied in (Kai et al. 2023; Shoji et al. 2020), the HYLON swirled burner (Capurso et al. 2023; Marragou et al. 2022) and the unconfined bluff-body burner developed at NTNU (Æsøy and Dawson 2023). The latter is particuarly appealing to study the coupling between differential diffusion and stretch, since it focuses on a fully-premixed lean hydrogen case and both strain and curvature effects are observed. Recent studies for the confined version of this configuration have emphasised the role of preferential diffusion on the ignition dynamics (Yahou et al. 2024), and the effects of stretch and heat losses in hydrogen blended flames (Kutkan et al. 2022b,a). Nevertheless, the coupled dynamics between differential diffusion and stretch, in particular its strain component, and its implication for flamelets-based approaches, remains unclear. The present work aims at investigating the NTNU unconfined test case using a flamelet-based approach coupled with a presumed filtered density function (FDF) closure, which was recently developed to incorporate effects of differential diffusion (Ferrante et al. 2024b) by extending the method proposed in (Mukundakumar et al. 2021) for laminar flamelet models to the context of LES with presumed FDF. This extended model was shown to be capable of capturing the fluctuations of local equivalence ratio at resolved level associated with the joint effect of differential diffusion and flame curvature, originating from turbulence and thermodiffusive instabilities. However, it remains unclear whether the effects of (tangential) strain can be captured within this framework. Since both strain and curvature produce macroscopic effects of leading order in premixed hydrogen flames, it is crucial to capture the interplay between differential diffusion and the effects of strain and curvature in the combustion modelling. This raises the question of whether an unstrained, one-dimensional flamelets database, as opposed to a database of strained flamelets, would be adequate for capturing the influence of strain on the flame dynamics. Despite the one-dimensionality of an unstretched flamelet suggests that this is not possible, the answer is not straightforward. At the macroscopic level, and similarly to a curved flame case, the thermochemical state of a strained flame is altered by the balance between convection and diffusion. Thus, a model for LES that can mimic this balance could in principle retrieve the correct thermochemical state starting from unstretched conditions. Studies in the context of hydrocarbon flames (Langella and Swaminathan 2016) have already shown that strained flamelets are not necessarily required if most of the strain is resolved, even in presence of local extinctions (Chen et al. 2020; Soli et al. 2021). In the context of hydrogen premixed flames Berger et al. (Berger et al. 2022) suggested that adding a mixture fraction among the controlling variables provides a greater improvement of thermochemical states parametrization in a turbulent (hydrogen) jet flame, than choosing curvature, stretch or strain. On the contrary, the *a priori* analysis conducted by Böttler et al. (Böttler et al. 2024) indicated that incorporating both strain and curvature variations in the flamelets database is necessary to capture all the thermochemical states. Nevertheless, the deviation of the reacting states predicted by the DNS from a flamelet database built only using unstretched flamelets at different mixture fraction was mostly associated to minor species, while smaller deviations were observed for main species, temperature and reaction rate. Berger et al. (Berger et al. 2025) recently developed and tested an LES flamelet model, incorporating preferential diffusion, based on unstretched premixed flamelets with varying equivalence ratio. Their results show that this manifold captures most of the effects of differential-diffusion and its coupling with flame front curvature (caused by thermodiffusive instabilities) and turbulence, and strain (induced by turbulence). In light of the above studies, it remains unclear whether in configurations characterised by intense strain the combined effects of strain, curvature and differential diffusion on the distribution of mixture fraction and reaction rate peak can be captured by the sole effect of resolved strain in an LES, and starting from a database of unstretched flamelets. In contrast with the slot-burner flame investigated in (Berger et al. 2025), the bluff-body stabilised flame analysed in the present study is dominated by relatively high mean strain-rate levels if compared to those of curvature. This offers an opportunity to assess the applicability of unstretched flamelet libraries in such regimes. In particular, the present work aims to: (i) validate the flamelet model for the current test case against available experimental data (PIV and OH*-chemiluminescence) and numerically investigate the flame characteristics; and (ii) evaluate the model's capability in capturing differential diffusion and strain effects on the flame's behaviour at the resolved level using a flamelet database built on unstrained flamelets.

## 2. Case study

The configuration studied in this work is the premixed bluff-body stabilised flame developed at the Norwegian University of Science and Technology (NTNU). This configuration is schematically illustrated in Fig. 1. The geometry is the same as investigated by (Æsøy et al. 2021), although the flame analysed in the present work is unconfined. Details on geometry and experimental data can be found on the TNF Workshop archive (Æsøy and Dawson 2023). The conical bluff body has a diameter $d_b = 13$ mm and half-cone angle $\alpha = 45°$. The hydrogen-air mixture is introduced with a temperature $T = 25°C$ through an annular duct with an external diameter $d_b = 19$ mm and a bluff-body holder diameter of 5 mm. The burner operates without confinement, allowing the flame to develop freely in an environment

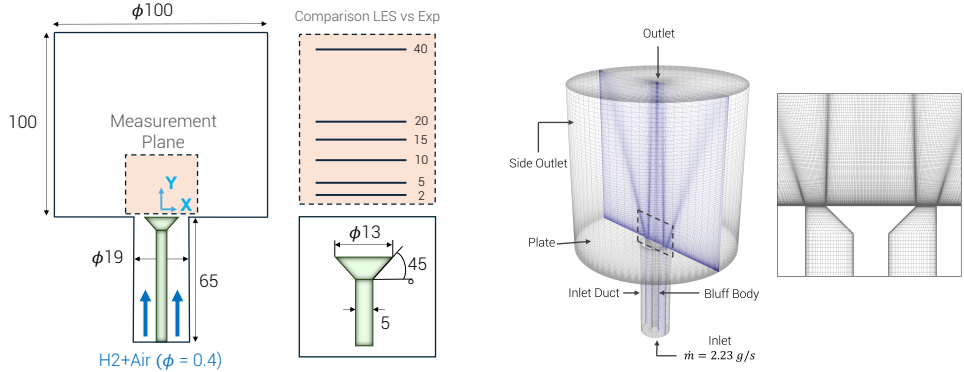

Figure 1.: Schematic of the NTNU premixed hydrogen burner (left), and corresponding numerical mesh and boundary conditions (right). All dimensions are reported in mm.

at atmospheric pressure and temperature.

The operating conditions are summarised in Tab. 1 together with laminar flame properties at the nominal equivalence ratio $\phi = 0.4$. Experimental dataset consists of OH* chemiluminescence and RMS and time-averaged velocity. these are measured at 2 mm, 5 mm, 10 mm, 15 mm and 20 mm from the bluff-body base in the stream direction for the cold flow experiment, and at 2 mm, 5 mm, 10 mm, 20 mm and 40 mm for the reactive flow case.

## 3. Methods

Large-eddy simulations with flamelet based thermochemistry and presumed FDF closure are employed in this work. The Favre-filtered, reactive Navier-Stokes include transport equations for continuity, momentum and absolute enthalpy (sum of formation and sensible enthalpies). Additional equations are solved for the combustion model. These equations will be discussed in the next section. The sub-grid scales (sgs) are modelled using a one-equation approach, involving an additional transport equation for the sgs turbulent kinetic energy $k_{sgs}$ (Yoshizawa 1986). Flamelet models rely on the assumption that turbulent eddies can stretch and wrinkle the flame, but cannot affect its internal structure due to the smaller characteristic time and length scales of chemistry with respect to turbulence. Turbulence and thermochemistry can thus be decoupled and the turbulent flame can be described by an ensemble of 1D laminar flames (flamelets), which are parametrised by a limited number of controlling variables. The sgs effect of turbulence on the reactions is modelled by presuming the shape of the controlling variables' subgrid distribution via a filtered density function (FDF). Effects of differential diffusion are included following the method proposed in (Mukundakumar et al. 2021) and extended for presumed FDF approaches as in (Ferrante et al. 2024b). Specific details of combustion and differential diffusion modelling are discussed next.

Table 1.: Operating conditions for the NTNU bluff-body burner.

| $\phi$ | $\dot{m}_{mix}$ [kg/s] | $P_{th}$ [kW] | $s_l$ [m/s] | $\delta_l$ [m] | $\tau_f$ [s] |
|---|---|---|---|---|---|
| 0.4 | $2.287 \times 10^{-3}$ | 3.5 | 0.21 | $6.5 \times 10^{-4}$ | 0.0031 |

### 3.1. Combustion Model

The flamelets database consists of premixed unstretched flamelets at different equivalence ratios, parametrised using a progress variable $c$ based on water mass fraction $Y_{H2O}$, and the Bilger's mixture fraction $z$ (Bilger et al. 1990). These are defined as:

$$c = \frac{Y_{H_2O}(z) - Y_{H_2O}^u(z)}{Y_{H_2O}^b(z) - Y_{H_2O}^u(z)}, \qquad z = \frac{\frac{1}{2W_H}(z_H - z_{H,o}) - \frac{1}{W_O}(z_O - z_{O,o})}{\frac{1}{2W_H}(z_{H,f} - z_{H,o}) - \frac{1}{W_O}(z_{O,f} - z_{O,o})}, \tag{1}$$

where superscripts $u$ and $b$ refer to unburnt and burnt mixtures respectively, subscripts $f$ and $o$ indicate fuel and oxidiser streams, and $z_i$ refers to the elemental mass fractions. Transport equations for the Favre-filtered controlling variables $\tilde{c}$ and $\tilde{z}$ are solved together with the corresponding variance equations $(\widetilde{\sigma_c^2}, \widetilde{\sigma_z^2})$ and the transport equation for the absolute enthalpy $\tilde{h}$. These equations read:

$$\bar{\rho}\frac{D\tilde{c}}{Dt} = \nabla \cdot \left[\left(\overline{\rho D_c} + \frac{\mu_t}{Sc_t}\right)\nabla\tilde{c}\right] + \bar{\dot{\omega}}_c + S_c \tag{2}$$

$$\bar{\rho}\frac{D\tilde{z}}{Dt} = \nabla \cdot \left[\left(\overline{\rho D_z} + \frac{\mu_t}{Sc_t}\right)\nabla\tilde{z}\right] + S_z \tag{3}$$

$$\bar{\rho}\frac{D\tilde{h}}{Dt} = \nabla \cdot \left[\left(\overline{\rho \alpha} + \frac{\mu_t}{Pr_t}\right)\nabla\tilde{h}\right] + S_h \tag{4}$$

$$\frac{D\bar{\rho}\widetilde{\sigma}_c}{Dt} = \frac{\partial}{\partial x_i}\left[\left(\overline{\rho D_c} + \frac{\mu_t}{Sc_t}\right)\frac{\partial \widetilde{\sigma}_c}{\partial x_i}\right] - 2\bar{\rho}\widetilde{\varepsilon}_c + 2\frac{\mu_t}{Sc_t}\left(\frac{\partial \tilde{c}}{\partial x_i}\frac{\partial \tilde{c}}{\partial x_i}\right) + 2(\overline{c\dot{\omega}_c} - \tilde{c}\widetilde{\dot{\omega}}_c) \tag{5}$$

$$\frac{D\bar{\rho}\widetilde{\sigma}_z}{Dt} = \frac{\partial}{\partial x_i}\left[\left(\overline{\rho D_z} + \frac{\mu_t}{Sc_t}\right)\frac{\partial \widetilde{\sigma}_z}{\partial x_i}\right] - 2\bar{\rho}\widetilde{\varepsilon}_z + 2\frac{\mu_t}{Sc_t}\left(\frac{\partial \tilde{z}}{\partial x_i}\frac{\partial \tilde{z}}{\partial x_i}\right) \tag{6}$$

where $\mu_t$ is the turbulent sgs viscosity defined by the turbulence model, and $Sc_t$ and $Pr_t$ are the turbulent Schimdt and Prandtl numbers, respectively, both set to a constant value of 0.4. In both variance equations, the first term in the RHS accounts for molecular and subgrid turbulent diffusion, the second term models the dissipation of subgrid variance, and the third captures variance production from the interaction between subgrid fluctuations and resolved gradients. The scalar dissipation rate of mixture fraction in Eq. (6) $\tilde{\chi}_z = \widetilde{D}\nabla\tilde{z}\cdot\nabla\tilde{z} + \widetilde{\varepsilon}_z$ combines resolved and modeled subgrid contributions, with $\widetilde{\varepsilon}_z = c_z(\nu_t/\Delta^2)\sigma_z$ ($c_z = 2$ (Pitsch 2006)), $\Delta = V^{1/3}$, and $V$ the local cell volume. The final term in Eq. (5) provides a chemical source (or sink) for subgrid variance, where the term $\overline{c\dot{\omega}_c}$ is computed a priori and tabulated together with the other thermochemical variables. Unlike the mixture fraction, the subgrid scalar dissipation rate of the progress variable $\widetilde{\varepsilon}_c$ requires a specialized model. Here, it is computed as (Chen et al. 2020; Langella et al. 2015):

$$\widetilde{\varepsilon}_c = \left[1 - \exp\left(\frac{-0.75\Delta}{\delta_{th}}\right)\right]\left[(2K_c - \tau C_4)\frac{s_L}{\delta_{th}} + C_3'\frac{\epsilon_k}{k}\right]\frac{\widetilde{\sigma}_c}{\beta_c} \tag{7}$$

where $s_L$, $\delta_{th}$, and $\tau$ denote the laminar flame speed, thermal flame thickness, and heat release factor, respectively, all extracted from the flamelet library. The remaining coefficients are non-dimensional constants; further details appear in (Chen et al. 2020; Langella et al. 2018). The modeling parameter $\beta_c$, which reflects influences of flame curvature, diffusion, and reaction rates, exhibits scale dependence. In this work, $\beta_c$ is determined dynamically following the approach of (Langella et al. 2015; Ferrante et al. 2024a). Two simulations are conducted, respectively with and without taking differential diffusion into account, which are referred respectively as $\text{Le}_k \neq 1$ and $\text{Le}_k = 1$ in the remainder of this paper for simplicity. The diffusion coefficients of the controlling variables $D_c$ and $D_z$ are taken equal to the thermal diffusivity $\alpha$ when unity Lewis number is assumed for every species. The source terms $S_c$, $S_z$ and $S_h$ are non-zero for the cases where differential diffusion is taken into account. These and the corresponding diffusion coefficients will be discussed in the next section. The filtered reaction rate $\overline{\dot{\omega}}_c$ in Eq. (2) is calculated as

$$\overline{\dot{\omega}}_c\left(\widetilde{c}, \widetilde{z}, \widetilde{\sigma_c^2}, \widetilde{\sigma_z^2}\right) = \overline{\rho} \int_0^1 \int_0^1 \frac{\dot{\omega}_c(c,z)}{\rho(c,z)} P(c,z; \widetilde{c}, \widetilde{z}, \widetilde{\sigma_c^2}, \widetilde{\sigma_z^2})\, dc\, dz, \qquad (8)$$

where the progress variable reaction rate, $\dot{\omega}_c(c,z)$, is taken from the 1D laminar flamelets database. The joint filtered density function $P(c,z)$ is assumed to be the product of Beta distributions both for $c$ and $z$, whose shape also depends on their sgs variances $\widetilde{\sigma_c^2}$ and $\widetilde{\sigma_z^2}$. These quantities are computed by resolving their respective transport equations as in (Langella and Swaminathan 2016; Chen et al. 2020; Ferrante et al. 2024a). The filtered temperature is computed from the equation $\widetilde{h} = \widetilde{\Delta h_f^0} + \int_{T_0}^{T} \widetilde{C}_p(T')\, dT'$, where $C_p$ is the mixture specific heat at constant pressure and $\Delta h_f^0$ is the mixture enthalpy of formation. The mixture density is computed via an ideal gas state equation. The Favre-filtered values for $C_p$, $\Delta h_f^0$, mixture molecular weight $W$ (needed for the state equation), as well as heat diffusivity $\alpha$, are also precomputed in a manner similar to Eq. (8).

### 3.2. Differential Diffusion Modelling

In the present study, the effects of differential diffusion are included at the thermochemistry level by employing a mixture averaged diffusion model in the solution of the 1D premixed flamelets. For every species an averaged diffusion coefficient with respect to the mixture is computed as:

$$D_k^M = \frac{1 - Y_k}{\sum_{j,k \neq j}^{N} X_j / D_{jk}}, \qquad (9)$$

where $D_{jk}$ is the binary diffusion coefficient of species $j$ in species $k$ and $X$ and $Y$ indicate respectively the molar and mass fraction. As a consequence of differential diffusion, mixture fraction and enthalpy are no longer constant along the flamelet. In the LES the transport equations of enthalpy and controlling variables must be corrected, to be consistent with the transport model adopted in the flamelet solution, implying that the terms $S_c$, $S_z$ and $S_h$ in Eqs. (2)-(4) are non-zero. The correction derived and validated for laminar flames in (Mukundakumar et al. 2021), and successively extended and validated for the presumed FDF LES framework in (Ferrante et al. 2024b), is used here. In this approach, the controlling variables are defined as a linear combination of the $N_s$ species mass fractions, for example:

$z = \sum_{k=1}^{N_s} \xi_k Y_k$. Two scalars $(\beta_c, \beta_z)$ are then computed from the 1D flamelets as an average of species mass fractions weighted by their mean Lewis number $Le_k$ and their contribution to the controlling variable $\xi_k$, for example: $\beta_z = \sum_{k=1}^{N_s-1}(\xi_k - \xi_{N_s})/\text{Le}_k Y_k$. These scalars are then pre-integrated consistently with Eq. (8). The filtered molecular diffusion term for the controlling variables is computed from the thermal diffusivity and the gradient of these scalars $\nabla \cdot (\bar{\rho} \widetilde{D} \nabla \widetilde{\beta}_i)$. Two additional scalars are used to correct the total enthalpy diffusion flux, which is written as $\nabla \cdot (\bar{\rho} \widetilde{D} \widetilde{\beta_{h1}} \nabla \widetilde{T} + \bar{\rho} \widetilde{D} \nabla \widetilde{\beta_{h2}})$, where $\beta_{h1}$ and $\beta_{h2}$ account respectively for local variations of specific heat and heat conductivity, and enthalpy redistribution, caused by differential diffusion along a flamelet. The new set of transport equations now reads:

$$\bar{\rho} \frac{D\widetilde{c}}{Dt} = \nabla \cdot \left( \frac{\mu_t}{Sc_t} \nabla \widetilde{c} \right) + \nabla \cdot \left( \bar{\rho} \widetilde{D} \nabla \widetilde{\beta}_c \right) + \bar{\dot{\omega}}_c \tag{10}$$

$$\bar{\rho} \frac{D\widetilde{z}}{Dt} = \nabla \cdot \left( \frac{\mu_t}{Sc_t} \nabla \widetilde{z} \right) + \nabla \cdot \left( \bar{\rho} \widetilde{D} \nabla \widetilde{\beta}_z \right) \tag{11}$$

$$\bar{\rho} \frac{D\widetilde{h}}{Dt} = \nabla \cdot \left( \frac{\mu_t}{Sc_t} \nabla \widetilde{h} \right) + \nabla \cdot \left( \bar{\rho} \widetilde{D} \widetilde{\beta}_{h_1} \nabla \widetilde{T} + \bar{\rho} \widetilde{D} \nabla \widetilde{\beta}_{h_2} \right) \tag{12}$$

$$\frac{D\bar{\rho}\widetilde{\sigma}_c}{Dt} = \frac{\partial}{\partial x_i} \left[ \left( \overline{\rho D_c} + \frac{\mu_t}{Sc_t} \right) \frac{\partial \widetilde{\sigma}_c}{\partial x_i} \right] - 2\bar{\rho}\widetilde{\varepsilon}_c + 2\frac{\mu_t}{Sc_t} \left( \frac{\partial \widetilde{c}}{\partial x_i} \frac{\partial \widetilde{c}}{\partial x_i} \right) + 2(\overline{c\dot{\omega}_c} - \widetilde{c}\bar{\dot{\omega}}_c) \tag{13}$$

$$\frac{D\bar{\rho}\widetilde{\sigma}_z}{Dt} = \frac{\partial}{\partial x_i} \left[ \left( \overline{\rho D_z} + \frac{\mu_t}{Sc_t} \right) \frac{\partial \widetilde{\sigma}_z}{\partial x_i} \right] - 2\bar{\rho}\widetilde{\varepsilon}_z + 2\frac{\mu_t}{Sc_t} \left( \frac{\partial \widetilde{z}}{\partial x_i} \frac{\partial \widetilde{z}}{\partial x_i} \right) \tag{14}$$

### 3.3. Numerical Setup

The computational domain begins downstream of the three cylinders located 65 mm upstream of the bluff-body base. The open ambient surrounding the bluff body is modeled as a cylinder with a diameter $d = 100$ mm and height $h = 100$ mm. The domain is discretised using 2.5 million hexahedral cells, and the boundary layer is refined to ensure a $y^+$ lower than one. Within the reaction zone, cell sizes range from a minimum of 35 $\mu$m to a maximum of 400 $\mu$m, with an average size of approximately 100 $\mu$m, which corresponds to an average of 6.5 numerical cells within the laminar flame thickness. The quality of this mesh was assessed using the Cèlik criterium (Celik et al. 2005) in order to resolve at least the 80% of the turbulent kinetic energy throughout the computational domain. Fig. 1 (right) shows a cut plane of the computational mesh and the boundary conditions applied in the simulations. The Favre-filtered Navier Stokes equations under low-Mach approximation are solved using the finite volume method in OpenFOAM-v9. A constant time step of $3 \times 10^{-7}$ s is used to ensure the Courant number remains below 0.5 in every cell of the computational domain. The Pressure Implicit with Splitting of Operators (PISO) algorithm is selected for the pressure-velocity coupling and an implicit Euler scheme is used for the temporal discretisation. The thermochemical database is built on a set of freely propagating premixed hydrogen-air 1D flamelets at atmospheric pressure, with reactants temperature of 300 K and the mixture fraction in the reactants spanning the flammability range $z \in [0.007,\ 0.17]$. The 1D flamelets solutions are computed using the one-dimensional solver CHEM1D and the Burke mechanism (Burke et al. 2012) (11 species and 21 reactions) is used as reaction mechanism. The 4D flamelets database needed for the FDF approach (see Eq. (8)) is discretised with $n_c = 100$ equispaced points between 0 and 1 for the progress variable space, and $n_{\sigma_c^2} = 50$ points between 0 and $\widetilde{c}(1-\widetilde{c})$ for its subgrid variance space. The filtered mixture fraction space spans the flammability range, and a total of $n_z = 98$ equispaced points is used. A total of $n_{\sigma_z^2} = 15$ points are used for the mixture fraction subgrid variance, spanning exponentially the range $[0, \widetilde{z}(1-\widetilde{z})]$. Boundary conditions are assigned as follows and highlighted in Fig. 1. At the domain inlet, a mass flow rate of $\dot{m} = 0.00229$ kg/s with a value of mixture fraction of $z = 0.0125$ is imposed, which corresponds to the nominal equivalence ratio of 0.4 used in the experiments. Zero normal velocity gradients and total pressure corresponding to the atmospheric pressure are prescribed at the outlet(s) of the computational domain. The simulation is initialised with a mixture fraction equal to zero (representing air at 300 K) throughout the entire domain. Once the flow has convected the mixture fraction field downstream the bluff-body, a patch with progress variable $c = 1$ is imposed near the bluff-body to ignite the mixture.

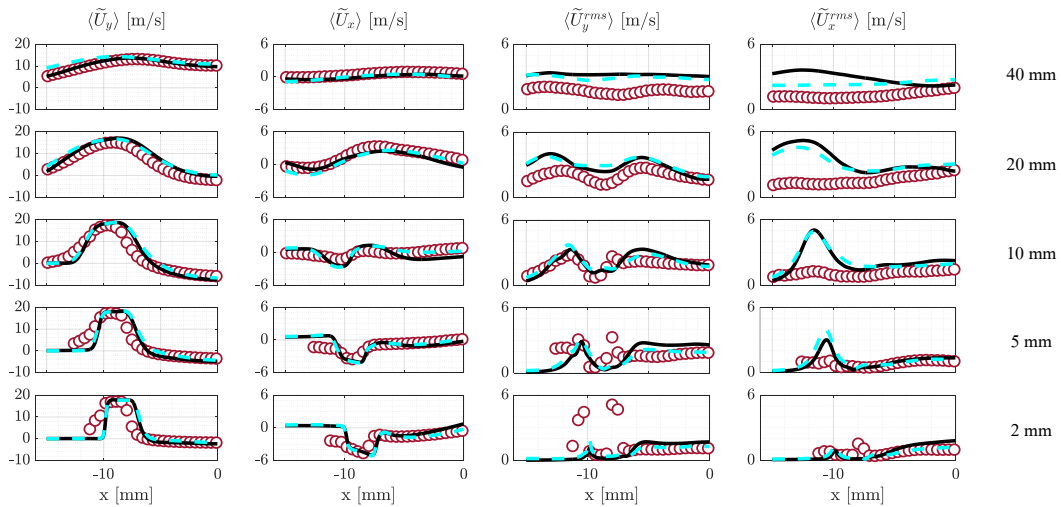

Figure 2.: Radial profiles of axial and radial velocity, and their RMS, at different axial locations. (LES with $Le_k = 1$: —, LES with $Le_k \neq 1$: ----, Exp ●)

## 4. Results and Discussion

### 4.1. Validation and flame structure

Comparisons between LES and experimental measurements of time-averaged and RMS axial and radial velocity components at various streamwise (axial) locations are shown in Fig. 2. Both the case with and without differential diffusion are reported, showing no significant differences in mean velocity flow field between the two models.

The simulations results align well with the experimental data, particularly in capturing the peaks of both velocity components and RMS fluctuations. However, some overestimation of the RMS of radial velocity is observed in the region of the shear layer around a height of 20 mm. This could be related to the fact that subgrid scales are affected by thermal dilatation effect in the flame normal direction, which is not taken into account in the estimation of these RMS. This argument is further suggested by the fact that the RMS of radial velocity was observed to better match the experimental measurements in the non-reacting case (not-shown). Also, comparisons of OH*-Chem measurements with time averaged reaction rate from LES, shown in Fig. 3, indicate that the LES accurately predicts both the shape and length of the flame. It can be further observed that the location of highest reaction rate moves towards the edge of the bluff-body when differential diffusion effects are included in the model, and the comparison with experimental data improves in this case.

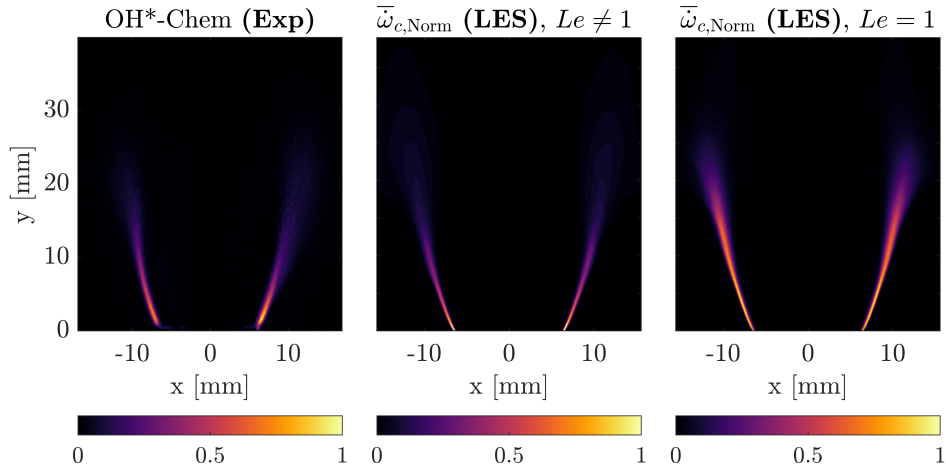

Figure 3.: OH*-Chemiluminescence from experiments (Æsøy and Dawson 2023) (left) and midplane contours of mean progress variable reaction rate obtained from LES results with (centre) and without (right) differential diffusion effects. All plots are normalised: experimental results have been normalised by their respective maximum OH*-Chem value, while the numerical results have been normalised by the maximum value of the $Le = 1$ case.

Further insight on the effects of differential diffusion is presented next. For an unstretched premixed flame, a decrease in mixture fraction is expected in the preheat zone ahead of the flame, followed by a re-increase in the flame region (Berger et al. 2022; Porcarelli and Langella 2024). This behaviour, categorised in the Introduction as a one-dimensional effect, is captured by the LES with differential diffusion model, as demonstrated in Fig. 4 by the presence of a region of mixture fraction with values around $\tilde{z} = 0.011$, which are lower than the nominal value $\tilde{z} = z_{nom} = 0.0125$. In fact, since the resolved curvature in this region is small and the positive resolved strain in the same region would increase the mixture fraction as compared to an unstrained case (Porcarelli and Langella 2024), the observed low values of $\tilde{z}$ necessarily imply that one-dimensional effects of differential diffusion are captured in the LES. Besides, the LES with differential diffusion model also predicts a significant increase in mixture fraction in the post flame region ($\tilde{c} > 0.5$), above its nominal value. This increase is only possible due to the combined effects of stretch and differential diffusion (positive strain and curvature), as documented in previous studies (Berger et al. 2022; Lapenna et al. 2021; Porcarelli and Langella 2024). This implies that the LES with unstretched flamelets thermochemistry can, at least qualitatively, mimic this interplay between resolved stretch and differential diffusion (multi-dimensional effect as categorised in the Introduction). As a consequence of this enrichment, the LES also predicts superadiabatic temperatures, with peaks around $\tilde{T} = 1800$ K observed in the region of the flame anchoring point (Fig. 4, right).

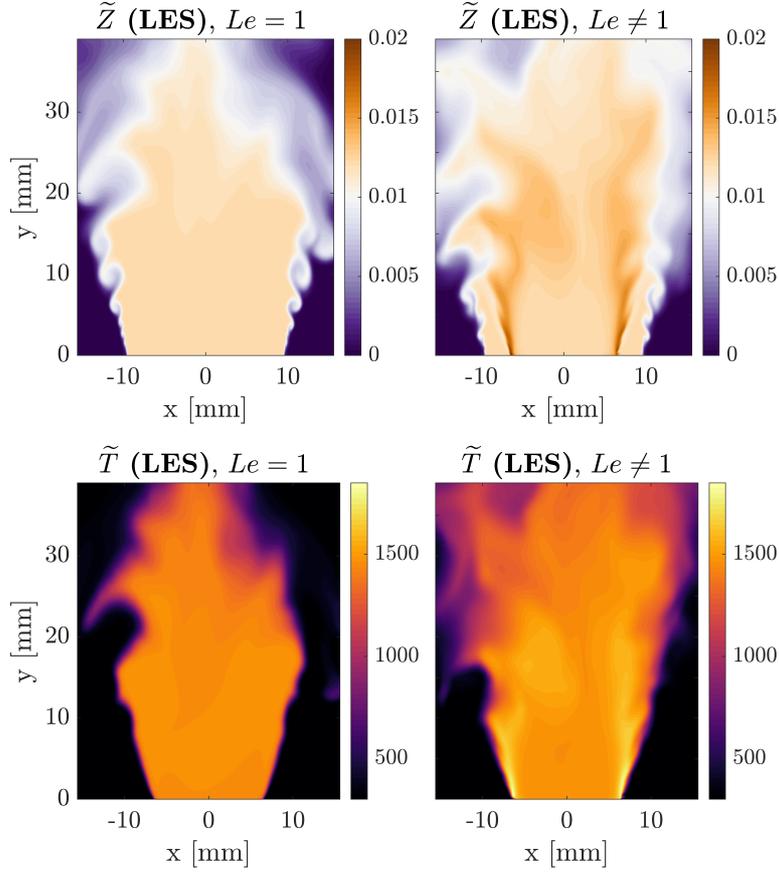

Figure 4.: Midplane contours of instantaneous filtered mixture fraction (left) and temperature (right) obtained from the LES without ($Le = 1$) and with ($Le \neq 1$) differential diffusion model. The mixture fraction field is the one obtained from its transport equation Eq. (11).

These peaks are found to correspond to values of progress variable $\tilde{c} > 0.85$. Consistently, a redistribution of the reaction rate field is predicted with respect to the unity Lewis number case, with higher values concentrated close to the flame anchoring point, also resulting in a shorter flame as observed with reference to Fig. 3. The above analysis highlights significant effects of differential diffusion in the studied configuration. Nevertheless, how the balance between strain and curvature is affecting the interplay between stretch and differential diffusion, and how the current modelling framework based on unstretched flames thermochemistry is capturing this interplay is yet unclear. Insights on this aspect are provided in the next section.

### 4.2. Average resolved stretch effects

In premixed flames, differential diffusion effects are tightly coupled with flame stretch $K$. This can be split into contributions of flame front curvature $k_f$ and flame tangential strain $S_t$ as

$$K = S_t + k_f s_d. \tag{15}$$

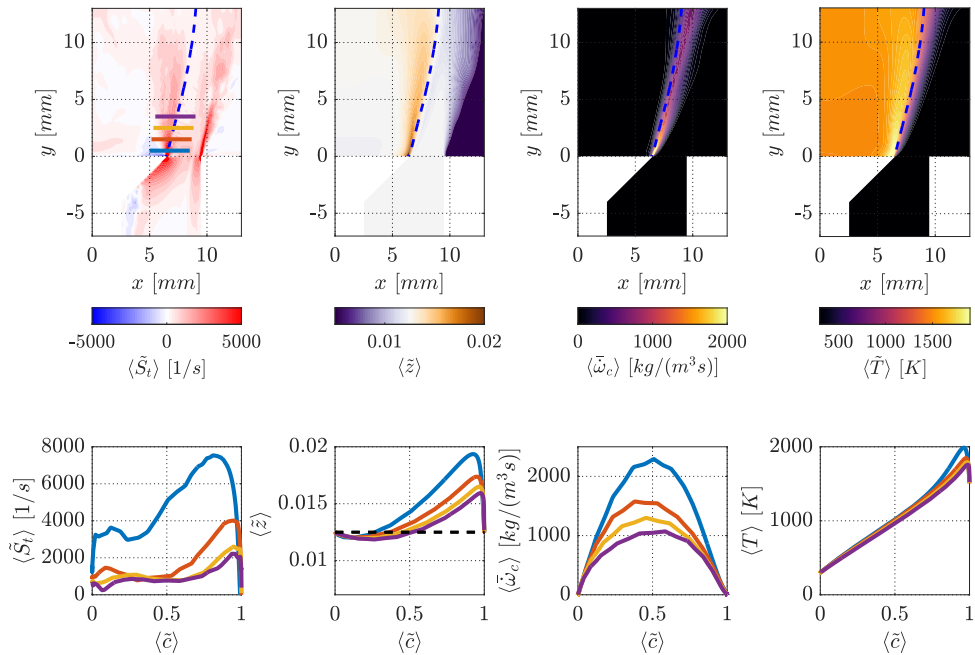

Figure 5.: Top row, from left to right: Midplane contours of mean resolved tangential strain rate $\langle \tilde{S}_t \rangle$, mean mixture fraction, $\langle \tilde{z} \rangle$, mean progress variable reaction rate, $\langle \bar{\tilde{\omega}}_c \rangle$, and mean temperature $\langle \tilde{T}_c \rangle$ obtained from LES with differential diffusion model. Isoline of mean progress variable $\langle \tilde{c} \rangle = 0.8$ is marked as a blue dashed line (----). Bottom row, from left to right: profiles of the same average quantities along the flame at the 4 axial locations marked in the A1 contour plot: $y = 0.5$ mm (—), $y = 1.5$ mm (—), $y = 2.5$ mm (—), $y = 3.5$ mm (—). The dashed line indicates the nominal mixture fraction value $z = 0.0125$.

In particular, flame curvature induced either by turbulence or by intrinsic thermodiffusive instabilities, is positively correlated with variations in mixture fraction, such that a convex curvature towards the reactants causes mixture enrichment upstream of the flame, while a cusp (concave towards reactants) produces a mixture fraction decrease (Williams 2018; Lipatnikov and Chomiak 2005). Similarly, positive flame tangential strain produces a mixture enrichment when coupled with differential diffusion (Porcarelli and Langella 2024). The contours in Fig. 5 indicate a region of intense positive tangential strain rate $\langle \tilde{S}_t \rangle \sim 7500$ 1/s being generated at the edge of the bluff body, where the flame anchors, which is due to the shear between the inlet flow and the recirculation region of hot gases.

The same figure reports radial profiles of mean flame tangential strain in progress variable space for different axial locations. As one would expect, tangential strain decreases downstream. Its peak location, found for mean progress variable values $\langle \tilde{c} \rangle > 0.8$, is also found to shift towards the product side as one moves downstream. The effect of mean resolved strain on mixture fraction is also visible in Fig. 5. Consistently with the study in (Porcarelli and Langella 2024), higher levels of strain observed near the bluff body base limit the (one-dimensional) differential diffusion effects on mixture fraction decrease in the low progress variable region, and enhance its re-increase for higher values of $\widetilde{c}$. This re-increase results in overshoot of mixture fraction above 50% in the region of higher strain, as compared to the nominal value. This in turns induces superadiabatic temperatures of the order of $\langle \widetilde{T} \rangle \sim 2000$ K. This phenomenon, which is linked to the relative magnitude of convective transport with respect to diffusive transport of mixture fraction due to differential diffusion in strained flames, is thus correctly captured by the LES at the resolved level. In the present configuration, unlike what was observed by Berger et al. (Berger et al. 2022), strain appears to be the main responsible for mixture fraction variations, dominating over curvature effects. The comparison with the present bluff-body reacting dynamics further suggests that thermodiffusive instabilities, which played a major role in the slot burner (Berger et al. 2022; Ferrante et al. 2024b) for a similar equivalence ratio and massflow rates, are somewhat suppressed by the significant strain at the bluff-body base, consistently with (Porcarelli and Langella 2024; Porcarelli et al. 2025). Figure 5 also indicates that not only the peak reaction rate, but also the extent of the area underneath the curve (which is proportional to the consumption speed in physical space), increases in correspondence of regions of higher tangential strain rates (see radial profiles). This is an expected effect of reacting mixtures at negative Markstein lengths (Law 1989), and suggests that capturing the correct redistribution of mixture fraction might be sufficient, in the context of flamelets combustion modelling, to predict the correct increase in reaction rates. This is further confirmed by the fact that the flame shape and length, thus the burning speed, is in very good agreement with the experimental observations (see Fig. 3), which is not the case without differential diffusion modelling. On the other hand, one has to keep in mind that strain effects on reaction rate are also linked to enthalpy and radicals redistribution through the flame front as indicated in the *a priori* analysis of (Böttler et al. 2024). In the latter, however, errors were observed to remain below 10% in terms of major species, temperature, density and reaction rates when using an unstretched flamelets database as compared to strained flamelets, with larger errors for minor species. Unfortunately, minor species mass fractions are not available in the experimental database for comparisons. Note that errors on minor species are commonly expected within flamelets based approaches even for hydrocarbon flames, where additional transport equations and methods are often used to improve the accuracy of their predictions (e.g. see (Van Oijen et al. 2016)). The above results suggests that effects of strain in lean hydrogen flames can be accurately predicted with unstretched flamelets thermochemistry, at least as long as most of the strain is resolved.

### 4.3. Reacting states

Figure 6 illustrates the joint probability density function of progress variable reaction rate $\overline{\dot{\omega}}_c$ and progress variable $\widetilde{c}$ at a random timestep. Only points on the flame with $\overline{\dot{\omega}}_c > 1$ are considered. Consistently with what observed by Berger et al. (Berger et al. 2022) for a jet flame at the same equivalence ratio, differential diffusion effects (left) result in reaction rate values higher than those found in an unstretched 1D laminar flame, which is due to the effect of stretch on local enrichment. Since these states are significantly more abundant than states associated to mixture fraction leaning, the global conditional average, $\langle \overline{\dot{\omega}} | \widetilde{c} \rangle$, is also significantly larger than the one associated to unstretched samples only (red line), highlighting the significant effect of stretch on the reaction rate. Note that the latter curve still presents values larger than those for the unstretched laminar 1D flamelet (black curve), due to diffusion of mixture fraction from the enriched pockets in the nearby regions of non-zero stretch. Figure 6 (left) also indicates that increasing levels of tangential strain (dashed lines) yield an increase of conditional reaction rate, in agreement with the trends observed in (Porcarelli and Langella 2024). These effects are completely absent when considering the case without differential diffusion (Figure 6, right), where the global conditional average matches the one obtained using unstretched samples only.

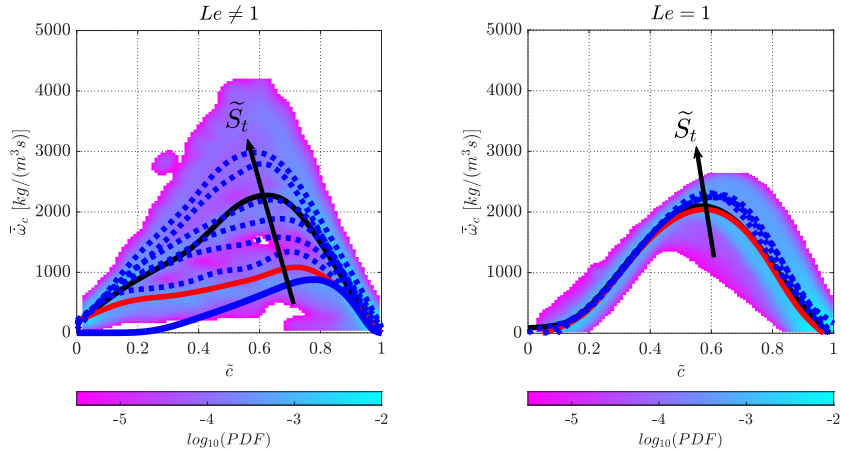

Figure 6.: Joint probability density function of progress variable reaction rate versus progress variable at a random timestep. Both cases with (left) and without (right) differential diffusion are shown. Solid lines represent global conditional average $\langle \bar{\dot{\omega}} | \widetilde{c} \rangle$ (—), conditional average restricted to unstretched reacting states $\langle \bar{\dot{\omega}} | \widetilde{c}, \widetilde{K} = 0 \rangle$ (—), and the reaction rate for an unstretched 1D premixed laminar flamelet at the same equivalence ratio (—). Dashed lines indicate the conditional averages at fixed levels of tangential strain $S_t$.

Also, as expected, increasing levels of resolved strain does not affect the conditional reaction rate. Figure 7 shows scatter plots of filtered mixture fraction $\widetilde{z}$ versus filtered progress variable $\widetilde{c}$, at a random timestep, coloured by mean stretch, strain and curvature respectively. Global conditional averages as well as conditional averages obtained only using samples at zero stretch, strain or curvature are also shown. Stretch and tangential strain are normalised by the chemical time scale $\tau_f$, and curvature is normalised by the laminar flame thickness $\delta_f$, whose values are reported in Tab. 1. The plot on the left indicates a strong correlation between positive values of stretch and mixture enrichment. The global conditional average of mixture fraction versus progress variable shows the typical profile of a strained premixed 1D flamelet in counterflow configuration (Porcarelli and Langella 2024), as already observed in Fig. 5, with an initial decrease in mixture fraction with respect to the nominal value followed by a reincrease and eventual overshoot. The conditional average of unstretched reacting state, $\langle \widetilde{z} | \widetilde{c}, \widetilde{K} = 0 \rangle$ (red line), is observed to align with the mixture fraction profile for an unstretched 1D premixed flamelet (blue line), confirming that stretch is driving this deviation. Consistently with Fig. 6 and the results in (Berger et al. 2022), the values on the "unstretched" conditional average curve are higher than those for the unstretched 1D flamelet. The two rightmost graphs in Fig. 7 further reveal that stretch effects are primarily driven by tangential strain in the present configuration, whereas curvature has a less pronounced influence. One can notice that the conditional average at zero curvature, $\langle \widetilde{z} | \widetilde{c}, \widetilde{\kappa}_f = 0 \rangle$ (red line, rightmost graph), aligns with the global conditional average (black line), consistently with results in (Berger et al. 2022). Furthermore, the conditional average at zero strain, $\langle \widetilde{z} | \widetilde{c}, \widetilde{S_t} = 0 \rangle$ (red line in the central graph), appears to align with the unstretched conditional mean (red line in the left graph). This milder influence of curvature effects, as compared to previous studies, is due to the fact that the positive tangential strain itself at the flame anchoring location might be limiting the onset of thermodiffusive instabilities (observed instead in (Berger et al. 2022)), thus limiting the occurrences of strongly curved flame fronts (Porcarelli and Langella 2024; Porcarelli et al. 2025). As a final remark, Fig. 7 (central graph) further illustrates the conditional mean at zero curvature for increasing strain levels. This confirms the argument for Fig.6 that increasing strain intensifies the local enrichment, as observed in (Porcarelli and Langella 2024) for laminar strained flamelets, leading in turn to the increase of reaction rate. To further characterise the influence of stretch, in Figure 8 the probability density functions of resolved stretch, tangetial strain and flame curvature, with and without differential diffusion, are shown. One can notice that curvature has a zero mode (although the mean remains slightly positive due to a mild skewness to the right), which is believed to be the result of only turbulent eddies motion. On the contrary, the PDF of strain indicates the presence of two peaks, which are believed in turn to correspond to the effect of turbulent eddies and to the applied strain in the shear region where the flame anchors (positive mode). This distribution is thus different to that observed in (Berger et al. 2022) where only a peak around zero strain was observed. It is interesting to note that the presence of differential diffusion causes an increase in the value of tangential strain associated with this second peak (strain itself intensifies), as well as an increased number of occurrences at this value. This is linked to the stabilization of the flame closer to the bluff-body edge characterised by higher strain levels and an overall shortening of the flame, and indicates that the hydrogen flame 'chases' the high strain region.

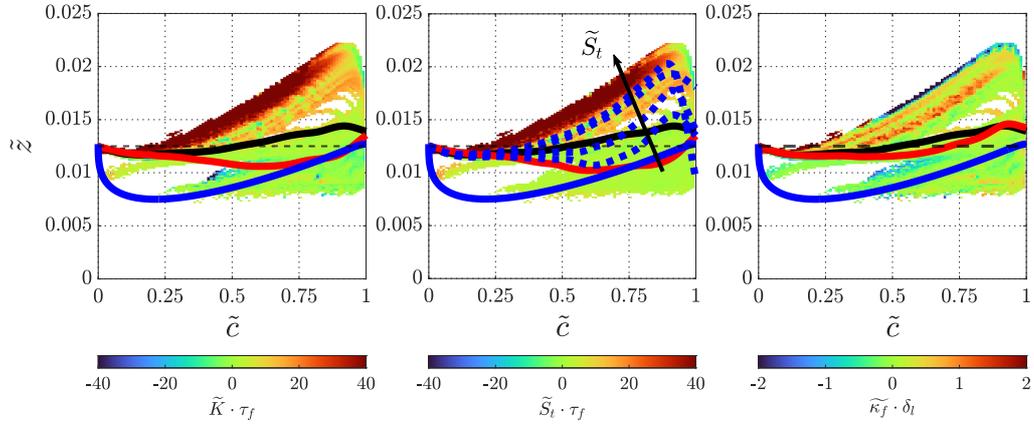

Figure 7.: Scatter plots of mixture fraction versus progress variable in the flame region. Colours represent mean resolved normalised stretch (left), strain rate (center), and curvature (right). Global conditional averages $\langle \widetilde{z} | \widetilde{c} \rangle$ are marked as solid black line (—). The nominal mixture fraction is marked with a black dashed line (⋯). The mixture fraction from 1D unstretched premixed flamelet at the same equivalence ratio is plotted with a solid blue line (—). Red lines (—) indicate conditional averages obtained considering points at zero stretch, strain and curvature, respectively. The dashed lines (⋯) represent increasing levels of strain rate.

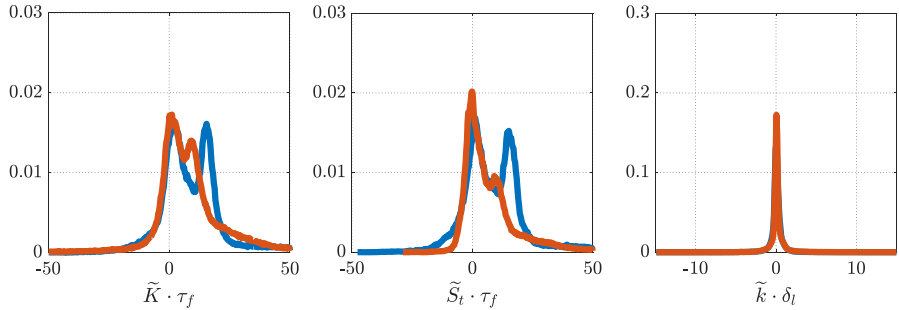

Figure 8.: Probability density function of resolved stretch, strain and curvature, showing the comparison between the unity Lewis number case (━) and with differential diffusion (━).

## 5. Conclusions

Large-eddy simulations with a flamelet-based combustion modeling and a recently-developed model to account for preferential diffusion, have been used to investigate the coupling between preferential diffusion and stretch in bluff-body stabilised lean premixed hydrogen flames subject to relatively high levels of flame tangential strain. The LES has been firstly validated against available experimental data of velocity and OH$^*$, showing that computational results are accurate. The preferential diffusion model was shown to be able to predict the local redistribution of mixture fraction in regions of non-zero strain and curvature. This redistribution was observed to cause the shift of the reaction source term peak towards the flame anchoring point, resulting in an increased flame-tangential strain and a shorter flame compared to the unity Lewis number case. Furthermore, the model captured the increase in reaction rate caused by positive strain by the sole use of unstretched flamelets thermochemistry. These findings demonstrate in the LES context that the considered differential diffusion model, based on an unstretched flamelets database, is effective in mimicking the peculiar physics of lean premixed hydrogen flames that exhibit coupling between stretch and differential diffusion, particularly in a configuration characterised by significant flame-tangential strain rate. This requires most of the strain and curvature to be resolved and the use of an appropriate correction for mixture fraction and progress variable. The insights provided within this work thus represent a step forward the definition of modelling strategies for highly-resolved LES simulations of lean premixed hydrogen flames.

**Acknowledgments**

AM and TG acknowledge the "e.INS - Ecosystem of Innovation for Next Generation Sardinia" funded by the Italian Ministry of University and Research under the Next-Generation EU Programme (National Recovery and Resilience Plan - PNRR, M4C2, INVESTMENT 1.5 - DD 1056 of 23/06/2022, ECS00000038, CUP F53C22000430001). GF and IL acknoweldge the Dutch Ministry of Economic Affairs and Climate under the TKI scheme and Safran SA for providing funding support to this project. IL further acknowledges financial support from the ERC Starting Grant OTHERWISE, grant n. 101078821. The authors further acknowledge use of the Delft Blue supercomputer to perform the simulations reported in the present work. This manuscript is made available under a CC-BY licence.

## Disclosure statement

The authors declare that they have no relevant financial or non-financial competing interests to report.

## References

Berger L, Attili A, Gauding M, Pitsch H. 2025. Les combustion model for premixed turbulent hydrogen flames with thermodiffusive instabilities: a priori and a posteriori analysis. Journal Fluid Mechanics. 1003:A33. https://doi.org/10.1017/jfm.2024.1201

Berger L, Attili A, Pitsch H. 2022. Synergistic interactions of thermodiffusive instabilities and turbulence in lean hydrogen flames. Combustion Flame. 244:112254. https://doi.org/10.1016/j.combustflame.2022.112254

Bilger R, Stårner S, Kee R. 1990. On reduced mechanisms for methane- air combustion in non-premixed flames. Combust. Flame. 80(2):135–149. https://doi.org/10.1016/0010-2180(90)90122-8

Böttler et al. 2024. Can flamelet manifolds capture the interactions of thermo-diffusive instabilities and turbulence in lean hydrogen flames? - an a-priori analysis. Int. J. Hydrog. Energy. 56:1397–1407. https://doi.org/10.1016/j.ijhydene.2023.12.193

Burke MP, Chaos M, Ju Y, Dryer FL, Klippenstein SJ. 2012. Comprehensive h2/o2 kinetic model for high-pressure combustion. International Journal Chemical Kinetics. 44(7):444–474. https://doi.org/10.1002/kin.20603

Cabra R et al. 2002. Simultaneous laser Raman-Rayleigh-LIF measurements and numerical modeling results of a lifted turbulent $H_2/N_2$ jet flame in a vitiated coflow. Proc. Combust. Inst. 29:1881–1888. https://doi.org/10.1016/S1540-7489(02)80228-0

Capurso T, Laera D, Riber E, Cuenot B. 2023. Nox pathways in lean partially premixed swirling h2-air turbulent flame. Combust. Flame. 248:112581. https://doi.org/10.1016/j.combustflame.2022.112581

Celik IB, Cehreli ZN, Yavuz I. 2005. Index of resolution quality for large eddy simulations. Journal Fluids Engineering. 127(5):949–958. https://doi.org/10.1115/1.1990201

Chen ZX, Langella I, Barlow RS, Swaminathan N. 2020. Prediction of local extinctions in piloted jet flames with inhomogeneous inlets using unstrained flamelets. Combust. Flame. 212:415–432. https://doi.org/10.1016/j.combustflame.2019.11.007

Ferrante G, Chen ZX, Langella I. 2024a. Dynamic modelling of subgrid scalar dissipation rate in premixed and partially premixed flames with differential filter. Appl. Therm. Eng. 248:123233. https://doi.org/10.1016/j.applthermaleng.2024.123233

Ferrante G, Eitelberg G, Langella I. 2024b. Differential diffusion modelling for les of premixed and partially premixed flames with presumed fdf. Combust. Theory Model.:1–36. https://doi.org/10.1080/13647830.2024.2389099

Kai R, Tokuoka T, Nagao J, Pillai AL, Kurose R. 2023. Les flamelet modeling of hydrogen combustion considering preferential diffusion effect. Int. J. Hydrog. Energy. 48(29):11086–11101. https://doi.org/10.1016/j.ijhydene.2022.12.164

Kutkan H et al. 2022a. Modeling of turbulent premixed ch4/h2/air flames including the influence of stretch and heat losses. J. Eng. Gas Turbine Power. 144:011020/1–011020/20. https://doi.org/10.1115/1.4051989

Kutkan H, Amato A, Campa G, Tay-Wo-Chong L, Æsøy E. 2022b. Les of turbulent premixed ch4/h2/air flames with stretch and heat loss for flame characteristics and dynamics. (Turbo

Expo; Vol Volume 3B: Combustion, Fuels, and Emissions). 06. p V03BT04A021. https://doi.org/10.1115/GT2022-82397

Langella I, Chen ZX, Swaminathan N, Sadasivuni SK. 2018. Large-eddy simulation of reacting flows in industrial gas turbine combustor. Journal Propulsion Power. 34(5):1269–1284. https://doi.org/https://doi.org/10.2514/1.B36842

Langella I, Swaminathan N. 2016. Unstrained and strained flamelets for les of premixed combustion. Combust. Theory Model. 20(3):410–440. https://doi.org/10.1080/13647830.2016.1140230

Langella I, Swaminathan N, Gao Y, Chakraborty N. 2015. Assessment of dynamic closure for premixed combustion large eddy simulation. Combustion Theory Modelling. 19(5):628–656. https://doi.org/https://doi.org/10.1080/13647830.2015.1080387

Lapenna PE, Lamioni R, Creta F. 2021. Subgrid modeling of intrinsic instabilities in premixed flame propagation. Proc. Combust. Inst. 38(2):2001–2011. https://doi.org/10.1016/j.proci.2020.06.192

Law C. 1989. Dynamics of stretched flames. Symposium (International) on Combustion. 22(1):1381–1402. https://doi.org/10.1016/S0082-0784(89)80149-3

Lipatnikov A, Chomiak J. 2005. Molecular transport effects on turbulent flame propagation and structure. Prog. Energy Combust. 31(1):1–73. https://doi.org/10.1016/j.pecs.2004.07.001

Marragou S, Magnes H, Poinsot T, Selle L, Schuller T. 2022. Stabilization regimes and pollutant emissions from a dual fuel ch4/h2 and dual swirl low nox burner. Int. J. Hydrog. Energy. 47(44):19275–19288. https://doi.org/10.1016/j.ijhydene.2022.04.033

Mukundakumar N, Efimov D, Beishuizen N, Van Oijen J. 2021. A new preferential diffusion model applied to FGM simulations of hydrogen flames. Combust. Theory Model. 25(7):1245–1267. https://doi.org/10.1080/13647830.2021.1970232

Pitsch H. 2006. Large-eddy simulation of turbulent combustion. Annu. Rev. Fluid Mech. 38(1):453–482. https://doi.org/https://doi.org/10.1146/annurev.fluid.38.050304.092133

Porcarelli A, Langella I. 2024. Mitigation of preferential diffusion effects by intensive strain in lean premixed hydrogen flamelets. Proc. Combus. Inst. 40(1-4):105728. https://doi.org/10.1016/j.proci.2024.105728

Porcarelli A, Lapenna PE, Creta F, Langella I. 2025. Stability analysis of thermodiffusively unstable counterflow lean premixed hydrogen flames. Proceedings Combustion Institute. 41:105906. https://doi.org/10.1016/j.proci.2025.105906

Regele JD, Knudsen E, Pitsch H, Blanquart G. 2013. A two-equation model for non-unity Lewis number differential diffusion in lean premixed laminar flames. Combust. Flame. 160(2):240–250. https://doi.org/10.1016/j.combustflame.2012.10.004

Shoji T et al. 2020. Effects of flame behaviors on combustion noise from lean-premixed hydrogen low-swirl flames. AIAA journal. 58(10):4505–4521. https://doi.org/10.2514/1.J059049

Soli A, Langella I, Chen ZX. 2021. Analysis of flame front breaks appearing in les of inhomogeneous jet flames using flamelets. Prog. Energy Combust. Sci. 49:189–208. https://doi.org/10.1007/s10494-021-00306-6

Van Oijen J, Donini A, Bastiaans R, ten Thije Boonkkamp J, De Goey L. 2016. State-of-the-art in premixed combustion modeling using flamelet generated manifolds. Progress Energy Combustion Science. 57:30–74. https://doi.org/10.1016/j.pecs.2016.07.001

Williams F. 2018. Combustion theory. CRC Press. https://books.google.it/books?id=2VNPDwAAQBAJ

Yahou T et al. 2024. The role of preferential diffusion on the ignition dynamics of lean premixed hydrogen flames. Proc. Combus. Inst. 40(1-4):105612. https://doi.org/10.1016/j.proci.2024.105612

Yoshizawa A. 1986. Statistical theory for compressible turbulent shear flows, with the application to subgrid modeling. Phys. Fluids. 29(7):2152–2164. https://doi.org/10.1063/1.865552

Æsøy E, Aguilar JG, Bothien MR, Worth NA, Dawson JR. 2021. Acoustic-convective interference in transfer functions of methane/hydrogen and pure hydrogen flames. J. Eng. Gas Turbine Power. 143:121017/1–121017/10. https://doi.org/10.1115/1.4051960

Æsøy E, Dawson JR. 2023. Tnf workshop. [https://tnfworkshop.org/bluff-body-hydrogen-flame-ntnu-trondheim/]. Accessed: 2024-11-27.